\documentclass[12pt, preprint]{aastex}

\newcommand{\etal}   {{\rm ~et al.}}
\newcommand{\kms}    {\ifmmode{{\rm ~km~s}^{-1}}\else{~km~s$^{-1}$}\fi}
\newcommand{\ta} {\tablenotemark{(a)}}
\newcommand{\tb} {\tablenotemark{(b)}}
\newcommand{\tc} {\tablenotemark{(c)}}
\newcommand{\td} {\tablenotemark{(d)}}
\newcommand{\te} {\tablenotemark{(e)}}
\newcommand{\tf} {\tablenotemark{(f)}}
\newcommand{\tg} {\tablenotemark{(g)}}

\shortauthors{Greenhill\etal}
\shorttitle{H$_2$O Maser Emission in AGN}

\begin{document}

\title{The Discovery of H$_2$O Maser Emission in 
Seven AGN and at High Velocities in the
Circinus Galaxy}

\author{
L. J. Greenhill,\altaffilmark{1}
P. T. Kondratko,\altaffilmark{1}
J. E. J. Lovell,\altaffilmark{2}
T. B. H. Kuiper,\altaffilmark{3}}
\author{
J. M. Moran,\altaffilmark{1}
D. L. Jauncey,\altaffilmark{2}
and G. P. Baines\altaffilmark{4}
}

\altaffiltext{1}{Harvard-Smithsonian Center for Astrophysics, 60 Garden St, 
Cambridge, MA 02138 USA; greenhill@cfa.harvard.edu, pkondrat@cfa.harvard.edu, moran@cfa.havard.edu.}

\altaffiltext{2}{Australia Telescope National Facility, CSIRO, Epping, NSW 2121
Australia; jlovell@atnf.csiro.au, djauncey@atnf.csiro.au.}

\altaffiltext{3}{Jet Propulsion Laboratory, 4800 Oak Grove Dr, Pasadena, CA 91109
USAi; kuiper@dsnra.jpl.nasa.gov.}

\altaffiltext{4}{BAE Systems, Canberra Deep Space Communication Complex (CDSCC), 
Paddy's River Rd, Tidbinbilla ACT, 2620, Australia; gbaines@anbe.cdscc.nasa.gov.}

\begin{abstract}

We report the discovery of H$_2$O maser emission at 1.35\,cm wavelength in
seven  active galactic nuclei (at distances up to $<80$ Mpc) during a survey
conducted at the 70-m diameter antenna of the NASA Deep Space Network near
Canberra, Australia.  The detection rate was $\sim 4\%$.  Two of the maser
sources are particularly interesting because they display satellite
high-velocity emission lines, which are a signature of  emission from the
accretion disks of supermassive black holes when seen edge on.  Three of the
masers are coincident, to within uncertainties of $0\rlap{.}''2$, with
continuum emission sources we observed at about $\lambda 1.3$~cm.  
We also report the discovery of
new spectral features in the Circinus galaxy H$_2$O maser that broaden the
known velocity range of emission therein by a factor of $\sim 1.7$. If the new
spectral features originate in the Circinus accretion disk, then molecular
material must survive at radii $\sim 3$ times smaller than had been believed
previously ($\sim 0.03$~pc or $\sim 2\times10^5$ Schwarzschild radii).

\end{abstract}


\section{Introduction}

Water maser emission ($\lambda 1.35$~cm in the rest frame) is known to trace
warm, dense gas at radii of 0.1 to 1 pc in the accretion disks surrounding
supermassive black holes in galactic nuclei \citep[e.g.,][]{mgh99}.  It can
also trace material heated by jet activity \citep[e.g.,][]{claussen98} and
wide-angle nuclear winds \citep{greenhill.circinus}. Emission from disks is
visible when they are viewed close to edge-on and amplification paths are
longest.  Several have been mapped with Very Long Baseline Interferometry
(VLBI): NGC\,4258 \citep{miyoshi95}, NGC\,1068 \citep{gg97}, and the Circinus
Galaxy \citep{greenhill.circinus}. ``Maser disks'' may also exist in IC\,2560
\citep{ishihara01}, NGC\,5793 \citep{hagiwara01}, and  Mrk\,1419
\citep{henkel02}, though confirmation awaits further study.

Triply peaked spectra characterize emission from accretion disks that are well
populated by masers.  Emission close to the systemic velocity of the host
galaxies (i.e., low-velocity emission) occurs where orbital motion is
transverse to the line of sight. High-velocity emission is symmetrically
offset by the disk orbital velocities and arises in regions where the disk
motion is parallel to the line of sight.   Velocities as high as $\sim
1100$\kms~have been observed (i.e., in NGC\,4258).

Water maser sources in active galactic nuclei (AGN) are important
astrophysical tracers in part because VLBI can provide maps of resolved disk
structure (e.g., warping) and dynamics (e.g., rotation curves and proper
motions), as in \citet{herrnstein99}. Unfortunately, only $\sim 30$ H$_2$O
masers are known in AGN, and few of these exhibit triply peaked spectra.  The
discovery of new masers is a priority and a challenge.  First, the emission is
typically weak, and surveys must invest substantial time observing each target
with the largest available apertures (e.g., 1 hour with a 100 m diameter
antenna). Second, mean detection rates in surveys are typically $\ll 10$\% for
Seyfert~II galaxies closer than $cz\sim 7000 $\kms~\citep{braatz97}.  Third,
because the orbital speeds of disks (and concomitant velocity range of maser 
emission) cannot
be known in advance, surveys must have instantaneous bandwidths of thousands
of \kms. Sufficiently broadband observing systems have become available to the
general community only recently.

We report the detection of seven new masers obtained in a high sensitivity,
broad bandwidth survey of 160 nearby AGN ($cz<8100$\kms).  The survey was
unusual for two reasons.  First, we used a 70-m antenna of the NASA Deep Space
Network (DSN) to achieve high  sensitivity \citep[see also][]{greenhill3735}. 
Second, we used a custom built, portable, 5350\kms-wide spectrometer.  The
survey, which is ongoing at DSN facilities in the Northern and Southern
Hemispheres, and the hardware will be discussed in detail elsewhere.
 
\section{Observations}

The 70-m DSN antenna located at Tidbinbilla, near Canberra, 
is equipped with a cooled 18 to 
26~GHz HEMT receiver. The left circular polarization channel is limited by a
selectable 600~MHz bandpass filter and downconverted to a band centered at
321.4~MHz.   A second downconversion and bandpass filter deliver a 400 MHz
baseband ($\sim 5350$\kms~at $\lambda 1.35$~cm) at the input of a 2-bit,
four-level, 4096-lag digital autocorrelator.

Survey observations were conducted between 2002 May and September.  The zenith
system temperature was typically 40\,K under good winter observing conditions,
and zenith opacities were typically $\sim 0.05$.  The temperature was
calibrated with respect to an ambient load.  The peak aperture efficiency 
was $48\pm5\%$, and we determined its dependence on elevation through
antenna temperature measurements of PKS\,1830$-$211 ($6.3\pm0.3$~Jy) and
PKS\,1921$-$293 ($10.9\pm0.4$~Jy).  We calibrated these flux densities against 
3C286 (2.6~Jy at 21.8~GHz) with the Australia Telescope Compact Array (ATCA)
\footnote{The Australia Telescope Compact Array is part of the Australia
Telescope, which is funded by the Commonwealth of Australia for operation as a
National Facility managed by CSIRO.} on 2002 September 9 and 12.  We measured
and corrected for pointing errors (at the 70-m antenna), which were under most
circumstances on the order of $4''$ or 8\% of the half-power beamwidth.
As a result, we estimate the flux density calibration of
spectra is uncertain by $\la 10\%$.

To construct total-power spectra for each AGN, we nodded the antenna between
signal and reference positions.  Switching times of 30 or 45 s were usually
sufficient to produce flat baselines in the spectra.  We removed residual
fluctuations by subtracting a running mean computed over intervals of 256
channels ($\sim 337$\kms).  Heliocentric velocities were computed from sky frequencies 
determined by the digital tuning of receiver elements, using
the radio definition of Doppler shift. This calibration should be accurate to
better than 0.1\kms. We  checked it by observing the H66$\alpha$ radio recombination
line in W\,33, which has a known Local Standard of Rest velocity of
$36.2\pm0.2$\kms~\citep{wbw79}, and for which we measured a 
velocity of $36.3\pm0.2$\kms.

\section{Detections of New Sources}

We observed 160 AGN with the Tidbinbilla antenna and obtained typical
noise levels of 10 to 20 mJy ($1\sigma$) with 30 minutes of on-source 
integration.  We discovered seven new H$_2$O masers (Table\,1, Figure\,1). Each detection
has been confirmed through observation on more than one day. We have
also measured positions for six of the new masers with the Very Large Array
(VLA) of the National Radio Astronomy Observatory (NRAO)\footnote{The National
Radio Astronomy Observatory is a facility of the U.S. National Science
Foundation operated under cooperative agreement by Associated Universities,
Inc.} or the ATCA. These observing tracks were also used to estimate 
continuum flux density levels from the underlying AGN at about
$\lambda 1$~cm (Table\,1). 
Where we detected continuum emission, it was coincident with the maser
emission, to within estimated uncertainties ($0\rlap{.}''2$).

Six masers lie in Seyfert\,II objects, and one lies in a nucleus whose
classification is ambiguous \citep[NGC\,2824;][]{veron01}. The masers in
NGC\,5643 and NGC\,6300 lie in galaxies that had been targeted in previous
searches. \citet{greenhill2002} reported noise levels of 59 to 87~mJy for
NGC\,5643 and 59~mJy for NGC\,6300 in $\sim 0.8$\kms~wide channels, after 
Hanning smoothing. \citet{braatz96} report noise levels of 119 and 109 mJy for the two  galaxies
respectively, with 0.8\kms~wide channels.  At the strength we measured
($\sim 300$~mJy in a 1.3\kms~channel without Hanning smoothing), the maser
emission in NGC\,5643 would have been detected at the 3 to $5\sigma$ level by
these earlier observations, from which we conclude the maser emission varies
significantly with time.  The emission in NGC\,6300 is sufficiently weak that
it would not have been detected by the earlier observations.

Two of the new maser sources exhibit discrete line complexes above and
below the systemic velocity, which probably correspond to 
the high-velocity emission
that is a characteristic of maser action in the  accretion disks of
supermassive black holes. ESO\,269-G012 displays red and blueshifted
high-velocity emission, symmetrically offset by $\sim 650$~km\,s$^{-1}$
from a narrow
line near the systemic velocity of the galaxy (Figure\,2).  
The inferred orbital speed is
exceeded only by that of NGC\,4258 among known maser hosts.  The high-velocity
line complexes extend over $\sim 100$\kms, which may be due to the radial
distribution of emitting gas.  If the  rotation curve is ``Keplerian''
($v\propto r^{-0.5}$), then the outer radius of the  molecular material in the
disk would be $\sim 1.4$ times the inner radius. NGC\,6926 exhibits 
high-velocity emission that is symmetrically offset by up to $\sim 
200$\kms~from 
the systemic velocity (Figure\,1). 
Because the blueshifted 
emission is relatively weak, we have used the VLA to confirm it is real.  
If the
masers in ESO\,269-G012 and NGC\,6926 lie in disks at radii of 0.1~pc, which
is the radius of the innermost masers in NGC\,4258, then the enclosed mass is
on the order of $10^6$ to $10^7$ M$_\odot$.  The corresponding centripetal 
acceleration
in the ESO\,269-G012 disk ($v^2/r$), which would be 
manifested by a secular velocity
drift in the low-velocity emission, is $\sim4$\kms\,yr$^{-1}$, which is large
enough to measure readily within one year.

The maser in NGC\,2824 lies in an early-type galaxy and displays an
anomalously broad line profile ($\sim 150$\kms~half power full width).  Broad
lines are also seen toward other early-type galaxies that host maser sources
(e.g., NGC\,1052, Mrk\,348), where the emission is seen toward radio jets
rather than accretion disks \citep{claussen98,peck01}.   However, any
correspondence between Hubble-type and maser characteristics is tentative. 
For example, the maser in IRAS\,F01063$-$8034 is broad \citep{greenhill2002},
but the galaxy is believed to be type Sa (though it is difficult to classify
because it is edge-on).  In NGC\,1386 and IRAS\,F22265$-$1826, the relative
positions of the masers and jets therein are unknown because the jets are
undetected at about $\lambda 1.3$~cm \citep{sbg99}.  For NGC\,2824, we have
established a 1 mJy ($5\sigma$) upper limit on continuum emission at $\lambda
1.3$~cm (Table\,1).

\section{New High-Velocity Gas in Circinus}

Previously reported H$_2$O maser emission in the Circinus galaxy comprises two
broad complexes of lines that bracket the systemic velocity \citep[$\sim
439$\kms;][]{freeman77}, with spectral features offset by as much as $\sim
200$\kms~ \citep{nakai95, braatz96,greenhillcircinusvar}.  VLBI images 
($\sim 15$~mJy
$1\sigma$ noise) resolve a warped accretion disk 
and show emission with somewhat larger $\sim \pm 260$\kms~Doppler shifts
\citep{greenhill.circinus, greenhill.circinus.03}.  This is the maximum 
detected orbital speed, and it
corresponds to a radius of $\sim 0.11$ pc.

We have detected weak ($\ll 0.1$ Jy) high-velocity emission that establishes a
$\sim 50$ to $\sim 900$\kms~range of emission (i.e., Doppler shifts up to
$\sim 460$\kms). The rotation curve of the Circinus accretion disk is
approximately Keplerian \citep{greenhill.circinus}.  If the new emission
arises in the accretion disk, then it lies at $\sim 0.03$ pc radius
($2\times10^5$ Schwarzschild radii for a $1.7\times10^6$ M$_\odot$ black
hole), and future VLBI studies will be able to map the rotation curve and
warp of the accretion disk over an expanded range of radii.

\section{Summary}

We have detected new H$_2$O maser sources in seven AGN, as well as new
high-velocity emission components in the Circinus galaxy.  Two of the new
masers exhibit high-velocity emission, indicative of emission from accretion
disks.  It should be possible to map these two disks with VLBI, trace their
rotation curves, and weigh the central engines that bind them.    Ultimately,
it may be possible to measure geometric distances to the two host galaxies,
which would be significant because both lie in the Hubble Flow.  One of the
new masers lies in an early-type galaxy, and we speculate that it might
therefore be associated with an as yet undetected jet.  We have also
discovered new high-velocity water maser emission in the Circinus galaxy, VLBI
observation of which could greatly improve our understanding of its accretion
disk.

\acknowledgements
We thank C. Timoc and Spaceborne, Inc. for designing and building the
correlator, and for technical support in the field.  We are grateful for the
invaluable support and assistance given by the management, operations, and
technical staff of the CDSCC, and in particular thank J. Lauf for his
technical assistance with our program early on. We also acknowledge the
contributions of C. Garcia-Miro (INTA), J. F. G\'omez,  E. Jim\'enez Bail\'on,
and I. de Gregorio-Monsalvo (LAEFF/INTA) to testing of the spectrometer and
observing software at the Madrid DSN station prior to the observing season in
Australia. This work was supported in part by R\&D funds of the Smithsonian
Astrophysical Observatory.

\begin{deluxetable}{l@{\extracolsep{-9pt}}ccrr@{\extracolsep{15pt}}r
@{\extracolsep{-4pt}}rc}
\tablewidth{6.8in}
\tablecaption{Newly Discovered H$_2$O Masers}
\tablehead{
     \colhead{Galaxy}               &
     \colhead{V$_{\rm sys}$\ta}     &
     \colhead{Date\tb}              &
     \colhead{$\alpha_{2000}$}      &
     \colhead{$\delta_{2000}$\tc}   &
     \colhead{$T$\td}               &
     \colhead{$1\sigma$ noise\te}   &
     \colhead{Continuum\tf}     \\
     \colhead{}             &
     \colhead{(\kms)}       &
     \colhead{(DOY)}        &
     \colhead{(hhmmss)}     &
     \colhead{(ddmmss)}     &
     \colhead{(s)}          &
     \colhead{(mJy)}        &
     \colhead{(mJy)}        
}

\startdata

NGC\,2824     & 2735  & 170 & 09~19~02.22 & $+$26~16~12.0 & 2370 & 12 & $<1$ \\

NGC\,2979     & 2695  & 152 & 09~43~08.65 & $-$10~23~00.0 & 4050 & 6.6& \nodata \\

NGC\,5643     & 1194  & 252 & 14~32~40.70 & $-$44~10~27.8 & 3840 & 6.8& 4.1 \\

NGC\,6300     & 1106  & 185 & \nodata     &   \nodata     & 3465 & 5.7& \nodata  \\

NGC\,6926     & 5851  & 183 & 20~33~06.11 & $-$02~01~38.9 & 5490 & 5.1& 2.9  \\

ESO\,269-G012 & 4868  & 152 & 12~56~40.51 & $-$46~55~34.4 & 4905 & 6.5& $0.6$ \\

IRAS\,F19370-0131&6060& 208 & 19~39~38.91 & $-$01~24~33.2 & 4995 & 5.8& $<2.8$\\

Circinus      & 439 & 168 & 14~13~09.95 & $-$65~20~21.2\tg & 13068 & 5.1& \nodata  \\

\enddata

\tablenotetext{(a)}{Heliocentric velocity, assuming the radio definition of
Doppler shift.}

\tablenotetext{(b)}{Discovery date. Expressed as day-of-year (2002).}

\tablenotetext{(c)}{Maser positions measured with the ATCA (ESO\,269-G012,
NGC\,5643) or VLA. Uncertainties are $\pm0\rlap{.}''2$. The masers lie within
the $1\sigma$ error circles associated with optical or infrared positions of
nuclei, except for NGC\,2979 where the maser-optical RA offset is
$+2\rlap{.}''3$.}

\tablenotetext{(d)}{Total integration time on-source.}

\tablenotetext{(e)} {Noise levels are for a $\sim 98$~kHz ($\sim 1.3$\kms)
channel.  The flux density scale is corrected for elevation dependence of
antenna gain and referenced to outside the atmosphere. }

\tablenotetext{(f)}{Continuum flux densities, measured with the VLA, at 22.435
and 22.485 GHz, or the ATCA, at 21.056 GHz. The upperlimits reflect five times the
RMS noise. The half-power beamwidths for the two arrays were $\sim
0\rlap{.}''4$ and $\sim 0\rlap{.}''8$, respectively, and the sources were
unresolved.}

\tablenotetext{(g)}{VLBI position \citep{greenhill.circinus}.}

\end{deluxetable}
 
\newpage

\begin{figure}[th]
\epsscale{0.8}
\plotone{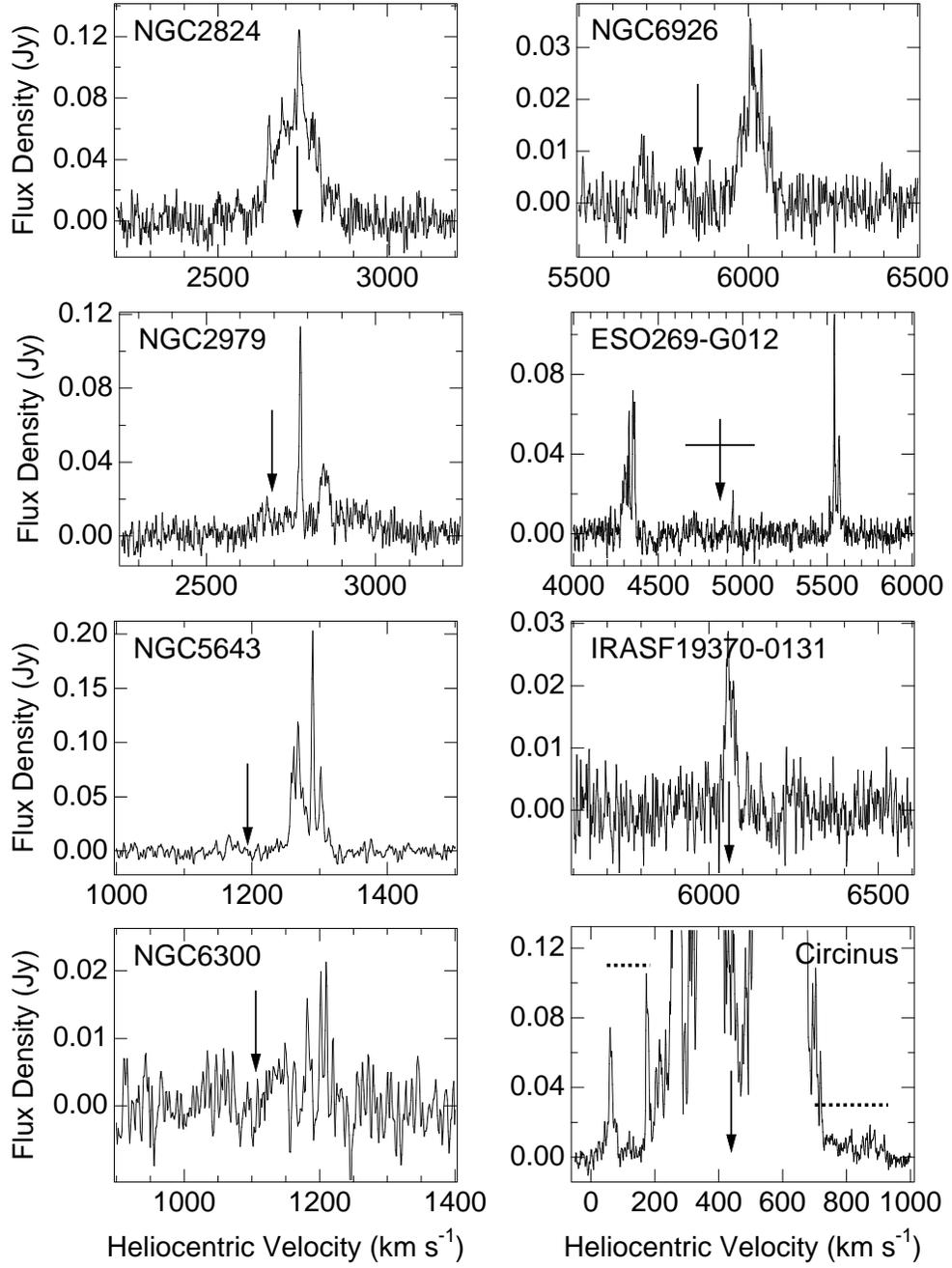}

\caption{Spectra of H$_2$O maser emission discovered with the 70-m DSN antenna
located at Tidbinbilla, near Canberra. Arrows show the systemic velocities of
the galaxies (assuming the radio definition of Doppler shift).  The systemic
velocity of ESO\,269-G012 is uncertain by 200\kms, which is indicated by the
horizontal error bar. The flux density scale for Circinus is expanded to show
the newly discovered emission, whose velocity range is indicated by the heavy
horizontal dashed lines. All spectra are Hanning smoothed to an effective
resolution of 2.6\kms. The spectra represent data collected on multiple days
that were separated by up to about one month.}

\end{figure}

\begin{figure}[th]
\epsscale{0.8}
\plotone{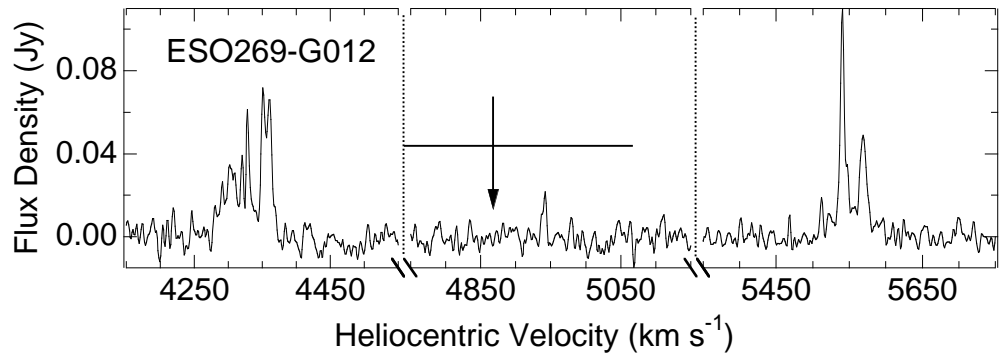}
\caption{Expanded spectrum of the H$_2$O maser emission in ESO\,269-G012.
Note that the velocity axis is not continuous.}
\end{figure}

\end{document}